\newcommand {\Mpc}   {\mbox{h$^{-1}$ Mpc \,}}

\newcommand {\ks}    {\mbox{km~s$^{-1}$\,}}

\newcommand {\sv}    {\mbox{$\sigma_{\mbox{{\small 
v}}}$\,}}

\documentstyle[psfig]{l-aa}
                     
\begin{document}

\thesaurus{
       (12.03.3;  
        12.04.3;  
        12.12.1;  
Universe
        11.03.1;  
        11.11.1)  
          }


\title{The ESO Nearby Abell Cluster Survey
       \thanks{Based on observations collected at the 
European Southern
               Observatory (La Silla, Chile)}
       \thanks{http://www.astrsp-mrs.fr/www/enacs.html}
      }

\subtitle{IV. The Fundamental Plane of clusters of galaxies}
\author{C.~Adami \inst{1}, A.~Mazure \inst{1} , A. Biviano \inst{2} , 
P. Katgert \inst{3} , G. Rhee \inst{4}}
\institute{IGRAP, Laboratoire d'Astronomie Spatiale, Marseille, France 
\and TeSRE, CNR, Bologna, Italy \and Sterrewacht Leiden, The Netherlands 
\and University of Nevada, Las Vegas, U.S.A. }                                    

\offprints{C.~Adami}
\date{Received date; accepted date}

\maketitle
\markboth{The ESO Nearby Abell Cluster Survey, IV. The 
Fundamental Plane of clusters of galaxies}{}

\begin{abstract}

We have used the ESO Nearby Abell Cluster Survey (ENACS) in
combination with the Cosmos Galaxy Catalogue, to investigate the 
existence of a Fundamental Plane (FP) for rich clusters of galaxies. 
The 20 clusters with the most regular projected galaxy distributions 
appear to define a quite narrow FP, which is similar to the FP found 
by Schaeffer et al., who used other clusters. Our cluster FP appears 
to be different from that of ellipticals, as well as from the virial 
prediction. The latter fact may have several physical explanations, 
or a combination thereof. If M/L varies with L this will change the 
FP slope away from the virial slope. Differences in dynamical 
structure between clusters will also produce deviations from the 
virial FP. In view of the long virialization time-scales in all but 
the very central parts of galaxy clusters, the deviation of the 
cluster FP from the virial expectation may also result from clusters 
not being totally virialized. The scatter of the observations around 
the cluster FP is fairly small. An important part of the observed 
scatter is likely to be intrinsic. If this intrinsic spread were due 
exclusively to deviations from the Hubble flow it would imply cluster 
peculiar velocities of at most about 1000 \ks.

\end{abstract}


\section{Introduction}
\label{s-intro}
One of the fundamental questions of modern cosmology is 
how the
gravitationally bound systems in our Universe have 
formed. Some clues
about the answer to this question may be provided by 
studying their
intrinsic properties, such as size, internal kinetic 
energy, and luminosity, as well as the 
relations between
these properties.

Elliptical galaxies are, e.g., known to populate a two 
dimensional
manifold in the ($R$,\sv,$L$)-space, referred to as the 
"fundamental
plane" (FP hereafter; Djorgovski \& Davis 1987, Dressler 
et al. 1987);
this relation between $R$, \sv and $L$ has been 
interpreted as arising
from the virial equilibrium of these systems. In virial 
equilibrium,
there is a relation between $R$, \sv, and $M$ (instead of 
$L$). In
order to understand the FP as arising from the virial 
relation, a
non-constant mass-to-light ratio is required for the 
ellipticals,
viz. $M$/$L$$ \propto M^{0.16}$ (e.g. Pahre et al. 1995).

The deviation from a constant $M$/$L$ ratio could be the
result of differences in the stellar population among 
ellipticals (e.g. 
Renzini \& Ciotti 1993), or of a partly 
dissipative formation
process (Capelato et al. 1995). The tilt of the FP 
relative to a
constant $M$/$L$ is also observed in the infra-red, where 
metallicity
effects are much reduced with respect to the optical 
(Pahre et
al. 1995). This may suggest that such a tilt is due to 
deviations from
homology in ellipticals as apparent in the correlations 
between
light-profile shape and r$_e$ or M$_B$ (as described by, 
e.g. Caon et
al. 1993 or Graham et al. 1996).

It has been shown through simulations that mergers of 
non-homologous
systems can produce a FP which slightly deviates from the 
expectation
from the virial condition (Capelato et al. 1995).  The 
fact that dwarf
ellipticals do not follow the same relation as regular 
ellipticals
(Bender et al. 1993) may then indicate a different 
formation process
for the two classes or, more simply, that interactions 
have a
different impact for large and small galaxies (Levine 
1996).

The FP is also important as a distance indicator, since 
\sv is a
distance-independent quantity, while $L$ and $R$ both 
depend on the
distance with different scaling laws. Before the FP was 
established,
the relation between \sv and $L$ found by Faber \& 
Jackson (1976) was
used as a secondary distance indicator for 
ellipticals
(similar to the relation discovered by Tully \& Fisher (1977)
for spiral galaxies).

Schaeffer et al. (1993) (hereafter S93) concluded on the 
basis of a
sample of 16 galaxy clusters that these systems also 
populate a
FP. They used a compilation by West, Dekel \& Oemler 
(1989) of
photometric data, and velocity dispersions for Abell 
clusters from
Struble \& Rood (1991). Schaeffer et al. also concluded that 
apparently there 
is a similar FP for all bound systems, which they supposed to span 
9 orders of magnitudes in
luminosity! This result was interpreted in the context of the
hierarchical structure formation scenario, as an indication
that globular clusters, galaxies and galaxy clusters have 
similar
formation processes. The dispersion in the FP should then 
reflect the
dispersion in the formation epoch.

There are several reasons for re-examining the existence and 
properties of a possible FP of clusters. First, the cluster sample 
used by S93 is rather heterogeneous: velocity
dispersions and interloper corrections have not been 
derived in the same 
way for all clusters, and photometric data come from 
about 20 different 
sources. Moreover, S93 use as characteristic scale the de 
Vaucouleurs
radius; however, as shown in Adami et al. (1998) 
(hereafter 
paper VII) the de Vaucouleurs profile gives a poor fit to 
the observed density profile of galaxy clusters. Indeed, 
we show in paper VII that this profile is too cusped within the 
central 100 kpc. 
For these reasons, the potentially important result of 
S93 needs 
confirmation. In the present paper, we re-examine the
evidence for a FP for galaxy clusters, using the large 
data-sample
from the ESO Nearby Abell Cluster Survey (ENACS, see 
Katgert et
al. 1996, hereafter paper I, Mazure et al. 1996, 
hereafter paper II,
Biviano et al. 1997, paper III, and Katgert et al. 1997, 
paper V), in
combination with the Cosmos Galaxy Catalogue. As discussed in paper V, 
our Cosmos data contain parts of the well-calibrated Edinburgh-Durham 
Southern Galaxy Catalogue (EDSGC, Heydon-Dumbleton et al. 1989), as 
well as somewhat less well calibrated parts of the Cosmos catalogue 
outside the EDSGC (courtesy of H. McGillivray). Although in paper V
we found evidence for a difference in the quality of the photometric
calibration between the two kinds of Cosmos data, there was no 
evidence for systematic magnitude offsets between the two parts of
the Cosmos Catalogue. So, for the present discussion we can
regard the ENACS and Cosmos datasets to be both homogeneous.

In \S~2 we give a short description of the sample of clusters that we
used in the present analysis, and of the two catalogues. In \S~3 we
discuss the methods with which we determined the core-radii, $R$, the
velocity dispersion, \sv, and the total cluster luminosities, $L$. In
\S 4 we derive the parameters that describe the FP of galaxy clusters
and in \S 5 we compare our results to other determinations of the FP
of galaxy clusters and early-type galaxies.

\section{The data sample}

\label{s-sample}

\begin{figure*}
\vbox
{\psfig{file=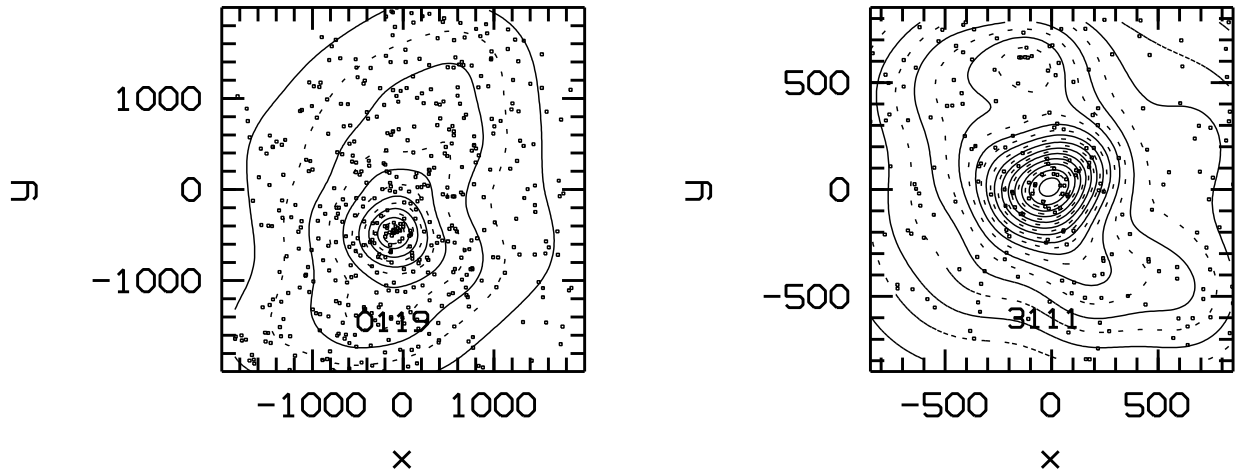,width=18.0cm,angle=270}}
\caption[]{Adaptive-Kernel maps of projected galaxy density for A0119 
and A3111, from the Cosmos catalogue.  Coordinates are in arcsec with 
respect to the geometric cluster center.}
\label{comp}
\end{figure*}

The current work is based on a subsample of 29 clusters from the
catalogue of Abell et al. (1989, ACO hereafter) with a redshift less
than 0.1, for which at least 10 ENACS galaxy redshifts are available
in the whole area. The typical redshift for these clusters is 0.07.
We have selected those clusters for which galaxy $b_j$ magnitudes and
positions were available to us from the Cosmos catalogue, and we have
limited the sample to clusters with very regular contours of projected
galaxy density.  In this way, we expect to reduce the scatter in the FP, 
due to the noise in the structural parameter measurements of clusters. 
In Fig. 1 we show surface density plots for two clusters (A0119 and 
A3111) which are representative for the total sample. For the other 27 
clusters, contour plots will be given in paper VII. 

We used the Cosmos data to calculate the integrated luminosity of the
cluster galaxies in a given selected area, as well as to determine
the characteristic scale of the cluster galaxy distribution. Note that 
the clusters A2734, A2764, A2799, A2800, A2911, A2923 and 
A3122 are in the EDSGC area around the Southern Galactic Pole, while 
the other 22 clusters are outside the EDSGC area but in the general 
Cosmos area. We assumed the entire Cosmos catalogue to have the same 
magnitude limit as the EDSGC subset, which is nearly complete at 
$b_j$ = 20.

The ENACS dataset was used to obtain an estimate of the degree of 
background (or foreground) field galaxy contamination in the cluster 
area. The ENACS data also allow us to determine the cluster velocity 
dispersions. However, we cannot calculate the integrated luminosity 
from those data. Spectroscopy was attempted for galaxy samples with 
well-defined completeness limits in magnitude. However, as the 
spectroscopy has not yielded redshifts for all galaxies that were 
observed, the galaxies with redshifts do not define a truly 
magnitude-limited sample. In paper V we illustrate this by comparing 
the differential $b_j$ magnitude distribution of the Cosmos galaxies 
to the differential $R_{25}$ magnitude distribution of the ENACS 
galaxies, for each cluster separately. Since the two catalogues may 
cover different regions of each cluster (typically, the ENACS samples 
are restricted to the inner regions and rarely extend beyond 1.5 \Mpc), 
we have selected only the ENACS and Cosmos galaxies in the 
intersection of both surveys.

Absolute magnitudes are derived from apparent magnitudes using the
cluster distance estimated from the mean cluster velocities, using a
Hubble constant $H_0$ = 100 km s$^{-1} \,$~Mpc$^{-1}$ and a
deceleration parameter $q_0=0$. We scaled the $R$$_{25}$ magnitudes 
to the $b_j$ system
by adopting a constant galaxy colour, $b_j$-$R$$_{25} = 1.5$ (see
paper V).  The Cosmos and ENACS magnitude distributions are quite
similar down to a given limiting magnitude which corresponds to the
({\em a priori} unknown) magnitude completeness limit of the ENACS
samples, which varies somewhat between clusters. If the Cosmos 
catalogue is complete to $b_j$ = 20, the ENACS catalogue is complete 
down to an absolute $R_{25}$ magnitude of about -20.

\section{Parameters of the Fundamental Plane}

\subsection{Characteristic Radii}
\label{ss-radius}

Using Cosmos galaxy positions, we have fitted four kinds of density
profiles to the galaxy distribution of each cluster. The background 
density is determined for each cluster and for each density profile. 
We used the profiles given by King (1962), by Hubble (e.g. Bruzual \& 
Spinrad 1978), by de~Vaucouleurs (1948) and by Navarro et al. (1996,
which we will refer to as the NFW profile). These profiles were 
generalized by allowing the exponents to vary. Each profile has its 
characteristic scale, which we determined by Maximum Likelihood 
fitting (see paper VII for details).  Although we used all four types 
of profile, it was found that only the King and Hubble profiles provide 
good fits to the observed clusters density profiles. The de~Vaucouleurs 
and NFW profiles have cusps and do not provide a good 
fit to the projected galaxy density in the central regions of clusters, 
which do not show a significant central cusp. Therefore, we mostly use 
the characteristic radii of the King and Hubble profiles in the present 
discussion; these are listed (with their errors) in table ~1.

As we need to adopt an area within which to define the cluster
properties, we chose to consider a square of 10 core radii size
centered on the cluster center assuming a King profile. The cluster
center we adopted was determined iteratively by fitting a 
King profile. The centers of the clusters used in the present analysis 
are given in table~1. The chosen area is expected to include a large 
fraction of the true ''physical cluster'', as the galaxy volume density 
at 5 King core radii is about 1~\% of the central density. For the other 
profiles the contrast between the central density and that at 5 King 
core radii is at least as large as for the King profile.

\subsection{Velocity dispersions}
\label{ss-disp}

\begin{table*}
\caption[]{Data for the sample of 29 clusters. Col.(1) lists the 
cluster name, col.(2) the velocity dispersion, cols.(3) and (4) the 
King (K) and Hubble (H) radius, cols (5) and (6) N$_{ENACS}$ and 
N$_{Cosmos}$, and cols (7) and (8) the fitted center of the clusters. 
In clusters marked by a $\dag$ interlopers have been removed with
the method of den Hartog $\&$ Katgert (1996).}
\begin{flushleft}
\small
\begin{tabular}{ccccrr@{\hspace{1.0cm}}rr}
\hline
\noalign{\smallskip}
Name & $\sigma $ (km s$^{-1}$) & $R$$_c$ K(kpc) & $R$$_c$ H(kpc) & 
N$_{ENACS}$ & N$_{Cosmos}$ & $\alpha$ (2000) & $\delta$ (2000) \\
\noalign{\smallskip}
\hline
\noalign{\smallskip}
A0013 & 1139$\pm 222$ & 68$\pm 18$ & 122$\pm 14$ & 12 & 57 & 
00:13:34.5 & -19:29:38 \\
A0087 & 510$\pm 107$ & 118$\pm 12$ & 263$\pm 51$ & 21 & 81 & 
00:43:00.7 & -09:50:37 \\
A0119 $\dag$ & 912$\pm 130$ & 55$\pm 14$ & 73$\pm 10$ & 27 & 76 & 
00:56:20.1 & -01:15:51 \\
A0151 $\dag$ & 911$\pm 159$ & 56$\pm 9$ & 94$\pm 16$ & 20 & 81 & 
01:08:50.1 &  -15:25:05 \\
A0168 $\dag$ & 549$\pm 37$ & 161$\pm 42$ & 178$\pm 28$ & 82 & 262 & 
01:15:14.9 & 00:15:23 \\
A0367 & 1096$\pm 145$ & 128$\pm 21$ & 162$\pm 26$ & 19 & 84 & 
02:36:35.2 & -19:22:16 \\
A0514 $\dag$ & 1187$\pm 107$ & 90$\pm 25$ & 117$\pm 33$ & 31 & 106 & 
04:48:15.1 & -20:27:26 \\
A1069 & 1002$\pm 96$ & 219$\pm 65$ & 411$\pm 65$ & 32 & 202 & 
10:39:47.9 & -08:40:46 \\
A2362 & 420$\pm 64$ & 110$\pm 32$ & 156$\pm 28$ & 17 & 86 & 
21:39:03.3 & -14:21:10 \\
A2480 & 805$\pm 190$ & 101$\pm 28$ & 272 & 10 & 68 & 
22:46:10.7 & -17:41:22\\
A2644 & 290$\pm 48$ & 76$\pm 22$ & 118$\pm 33$ & 12 & 151 &  
23:41:02.0 & 00:05:30 \\
A2734 $\dag$ & 648$\pm 66$ & 105 $\pm 22$ & 137 & 43 & 134 &  
00:11:22.9 & -28:50:55 \\
A2764 & 949$\pm 122$ & 101$\pm 22$ & 126$\pm 25$ & 14 & 133 & 
00:20:29.5 & -49:14:14 \\
A2799 & 427$\pm 90$ & 46$\pm 11$ & 48$\pm 15$ & 11 & 28 & 
00:37:24.1 & -39:09:01 \\
A2800 & 461$\pm 49$ & 98$\pm 25$ & 87$\pm 26$ & 19 & 71 & 
00:37:58.7 & -25:05:17 \\
A2854 & 328$\pm 118$ & 66$\pm 17$ & 70$\pm 11$ & 10 & 49 & 
01:00:47.2 & -50:32:38 \\
A2911 & 513$\pm 121$ & 109$\pm 28$ & & 21 & 130 & 
01:26:08.0 & -37:57:26 \\
A2923 & 401$\pm 87$ & 142$\pm 19$ & 96$\pm 19$ & 14 & 47 & 
01:32:28.8 & -31:04:41 \\
A3111 & 761$\pm 281$ & 99$\pm 26$ & 180$\pm 23$ & 22 & 98 & 
03:17:49.0 & -45:43:38 \\
A3112 $\dag$ & 865$\pm 222$ & 229$\pm 29$ & 433$\pm 65$ & 69 & 403 & 
03:17:58.5 & -44:14:21 \\
A3122 $\dag$ & 898$\pm 76$ & 149$\pm 20$ & 319 & 53 & 218 & 
03:22:14.0 & -41:19:14 \\
A3128 $\dag$ & 895$\pm 60$ & 362$\pm 29$ & 520$\pm 39$ & 123 & 867 & 
03:30:37.9 & -52:31:51 \\
A3141 & 758$\pm 124$ & 176$\pm 51$ & & 13 & 92 & 
03:36:54.3 & -28:04:20 \\
A3158 $\dag$ & 1310$\pm 123$ & 101$\pm 15$ & 104$\pm 12$ & 35 & 205 & 
03:43:04.6 & -53:38:40 \\
A3202 & 485$\pm 82$ & 64$\pm 20$ & 132$\pm 39$ & 12 & 51 & 
04:00:55.5 & -53:41:17 \\
A3733 & 768$\pm 249$ & 37$\pm 11$ & 84$\pm 15$ & 10 & 44 & 
21:01:34.7 & -28:02:42 \\
A3764 & 829$\pm 130$ & 60$\pm 16$ & 97$\pm 12$ & 12 & 45 & 
21:25:47.3 & -34:42:44 \\
A3825 $\dag$ & 947$\pm 87$ & 108$\pm 16$ & 227$\pm 50$ & 27 & 141 & 
21:58:26.0 & -60:22:03 \\
A3827 & 962$\pm 407$ & 57$\pm 29$ & 102$\pm 13$ & 11 & 99 & 
22:01:52.0 & -59:56:42 \\
\hline	   
\normalsize
\end{tabular}
\end{flushleft}
\label{t-data1}
\end{table*}

The velocity dispersions were obtained from the ENACS data-base. All 
29 clusters considered here have more than 10 galaxies in the
selected area and in the main group identified in radial velocity
space (see paper I). While in paper I the groups were defined using a
fixed gap of 1000 \ks in radial velocity space, here we slightly
modify this criterion to account for the fact that a fixed gap can
overestimate the number of groups when the number density of galaxies
is too low (simply because it is more likely to find larger gaps in
sparse data-sets). The new ''variable'' gap, which we will call {\em 
density gap,} is based on simulations of the occurrence of gaps of a 
given size in distributions of varying number of objects, drawn from 
the same Gaussian distribution. The density gap follows from the 
expression: 500 (1 + exp(-(n-6)/33)) km s$^{-1},$ where $n$ is the
number of galaxies in the redshift survey of a given cluster.

We stress that the criterion used in paper I works very well for the
ENACS datasets, since the number of galaxies does not vary too much. 
However, when we consider cluster datasets drawn from the literature, 
large differences in the number densities may occur. This is the case 
when one includes, e.g., the Virgo or the Coma clusters, for which 
redshifts are known for more than 500 galaxies. In these cases a
fixed gap-size fails to identify the main cluster structure, and
merges systems which are likely to be separate entities. For the present
paper, we could have maintained the fixed gap definition, but since we 
will in the future also include large datasets such as that of Coma, 
we prefer to use the density gap already in this discussion. We stress 
that using this gap definition, the membership of the ENACS main systems 
hardly changes, compared to paper I. 

Similarly to what we did in paper II, we identified interlopers in 
the systems by using both the spatial and the velocity information. 
More specifically, we applied the technique developed by den Hartog and 
Katgert (1996) to clusters with at least 50 galaxies left after the 
gapping in velocity space. These clusters are marked in table 1. The 
effect of the interloper removal was discussed at length in paper II. 
For systems with less than 50 galaxies the method becomes unreliable 
so we have not applied interloper rejection to these systems.

For the clusters thus defined, we calculate the velocity dispersion by
a biweight technique using the ROSTAT package (Beers et al.  1990),
which is ideally suited for poorly sampled and/or non-Gaussian 
distributions. As it is difficult to say whether our velocity
distributions are truly Gaussian (in some cases we only have 10
galaxies), a classical velocity dispersion estimator might not give a
reliable value. Lax (1985) has shown, from simulations, that the
biweight estimator gives better results in that case. Errors were 
estimated from 1000 bootstrap resamplings for each cluster (see Stein 
1996).

The reliability of our velocity dispersion values can be checked by a
comparison with previous investigations. In paper II we obtained
velocity dispersions by using a fixed gap criterion, and using all
galaxies in the ENACS regions. The present velocity dispersion 
estimates were correlated with those in paper II, and the best-fit 
straight line has a slope of 0.98$\pm 0.15$ and an offset of 
113$\pm 130$, clearly consistent with a slope of 1 and an offset of 0.
Comparing the present velocity dispersions with those of Fadda et al. 
(1996), who used a combination of ENACS and literature data, we find a 
best-fit line with a slope of 0.90$\pm0.17$ and an offset of 235$\pm 
203$, which again is consistent with the hypothesis that the two 
estimates are equivalent.

\subsection{Luminosity}
\label{ss-lumi}

In order to determine the cluster luminosities, we used the Cosmos 
data, and followed the procedure described below.

\begin{enumerate}
\item  We summed up the individual apparent luminosities of all 
galaxies in the selected area, after K-correcting the Cosmos $b_j$ 
magnitudes following Frei \& Gunn (1994) and assuming that all 
galaxies lie at the average redshift of the main system (as determined
from the ENACS data).  We then converted from apparent to absolute
magnitudes (using the standard cosmological formulae, see, e.g. Lang
1980), assuming an M$_{B{_j}}$ solar magnitude of 5.53 (Lang, 1980,
Gullixson et al. 1995). The result is $L$$_{Cosmos}.$

\item We applied a correction to the integrated luminosities for the 
contamination by fore- and background galaxies, by making the
assumption that the fraction $C_1$=$1 - (L_{fore- + background}) /
{L_{tot}}$ of the luminosity of cluster galaxies in the selected field, 
is the same in the Cosmos sample as it is in the ENACS sample, i.e 
assuming that $C_1$=$C_{Cosmos}$=$C_{ENACS}$. This is only approximately 
true, because this fraction changes with limiting magnitude and the 
Cosmos limit is 0.5 to 1.0 magnitude fainter than the ENACS limit. 
However, for some clusters that were very well sampled in the ENACS data
we have verified that $C_1$ is not significantly biased.

This allows us to take advantage of the fact that all ENACS galaxies
have a measured redshift, so that membership assignment is relatively
straightforward (while of course many Cosmos galaxies do not have such
information available).  We can then separate the luminosity of the 
cluster members from that of fore- and background galaxies. We calculate 
the value of $C_{ENACS}$ for each cluster, which we assume to be equal to
$C_1$. We find the mean value of $C_1$ = 0.83 if we consider only 
clusters with more than 15 galaxies. For the clusters with relatively 
poor statistics, the individual corrections are not considered reliable. 
Therefore, we apply the mean correction to all clusters by taking the 
mean value $\langle$ $C_1$ $\rangle $= 0.83; this introduces an 
uncertainty in the luminosities of the order of the dispersion in the 
$C_1$ values, which is 0.10. We thus have:

$L_1 = \langle C_1 \rangle \times L_{Cosmos} $  

\item We correct for the incompleteness effect due to the Cosmos
magnitude limit as follows. We adopt a completeness magnitude for the
Cosmos catalogue of $b_j=20$. Following Ellis et al. (1996), we take 
into account all fainter cluster members by adopting the following 
Schechter (1976) luminosity function:

$\varphi (L)=\left( \frac L{L^{*}}\right) ^{\alpha _{S}}
e^{-\frac L{L^{*}}}$ 

with $\alpha _{S}=-1.1$, M$_{b_j}^{*}=-18.8$ and $L$ the
absolute luminosity.

The fraction of the total luminosity that we have 
observed is then equal to:

$F$=${\frac {\int_{L_{lim}}^{+\infty}L\varphi(L)dL} 
{\int_0^{+\infty}L \varphi(L)dL} }$ 

where the value of $L_{lim}$ follows for each cluster from the median
redshift and $b_{j,lim}$ = 20. Finally we have:

$L_2 = C_2 \times L_1  =  L_1 / F $.

We have checked the dependence of $C_2$ on $\alpha _{S}$. For the same
limiting magnitude and a redshift equal to 0.07 (the typical distance
of our clusters), we have a variation of less than 10\% of $C_2$ when
$\alpha$ is changed from 0.5 to 1.3. There is thus only a weak
dependence of the total luminosity on $\alpha$ when this parameter is 
in the commonly employed range of values (e.g. Carlberg et al. 1996, 
Lumsden et al. 1997).

\item There is an additional source of incompleteness at bright
luminosities. In fact, we have found that in some cases, bright ENACS 
galaxies have no Cosmos counterpart (see paper V). In those cases, we 
simply add $C_3$, i.e. the sum of the ENACS luminosity (scaled to the 
$b_j$ system) of the galaxies which inadvertently were not included in 
$L_{Cosmos}$:

$L_3$ = $L_2$ + $C_3$.

\item Finally, we correct for galactic extinction $A_{b_j}$ by using 
the Burstein $\&$ Heiles map (1982). We use the formula $A_{b_j}=4.25 
\times E(B-V)$. The correction to the luminosity is $C_4$ = 
10$^{0.4{\times}A_{b_j}}$. We have:
 
$L_{true}$ = $L_3$ $\times$ $C_4$.

\end{enumerate}

\begin{table}
\caption[]{Luminosities (in units of h$^{-2}$$L$$_{{b_j},\odot}$) and 
corrections for the 29 clusters.}
\begin{flushleft}
\small
\begin{tabular}{crcrrr}
\hline
\noalign{\smallskip}
Name & $L_{true}$ & $C_1$ & $C_2$ & $C_3$ & $C_4$ \\
\noalign{\smallskip}
\hline
\noalign{\smallskip}
A0013 & 5.90 10$^{11}$ & 1.00 & 1.35 & 0. & 1.00 \\
A0087 & 2.48 10$^{11}$ & 0.70 & 1.12 & 1.7 10$^{10}$ & 1.10 \\
A0119 & 3.14 10$^{11}$ & 0.86 & 1.08 & 8.8 10$^{10}$ & 1.12 \\
A0151 & 3.82 10$^{11}$ & 1.00 & 1.11 & 1.6 10$^{10}$ & 1.00 \\
A0168 & 6.20 10$^{11}$ & 1.00 & 1.08 & 6.9 10$^{10}$ & 1.00 \\
A0367 & 7.90 10$^{11}$ & 1.00 & 1.32 & 0. & 1.06 \\
A0514 & 5.09 10$^{11}$ & 0.85 & 1.20 & 2.5 10$^{10}$ & 1.04 \\
A1069 & 9.27 10$^{11}$ & 0.95 & 1.16 & 0. & 1.08 \\
A2362 & 2.68 10$^{11}$ & 0.60 & 1.15 & 1.9 10$^{10}$ & 1.12 \\
A2480 & 3.46 10$^{11}$ & 0.74 & 1.20 & 0. & 1.09 \\
A2644 & 5.72 10$^{11}$ & 1.00 & 1.19 & 1.5 10$^{10}$ & 1.09 \\
A2734 & 5.31 10$^{11}$ & 0.74 & 1.15 & 9.7 10$^{10}$ & 1.00 \\
A2764 & 7.69 10$^{11}$ & 0.77 & 1.19 & 0. & 1.00 \\
A2799 & 1.30 10$^{11}$ & 1.00 & 1.15 & 6.0 10$^{10}$ & 1.00 \\
A2800 & 3.35 10$^{11}$ & 0.74 & 1.16 & 1.9 10$^{10}$ & 1.00 \\
A2854 & 2.19 10$^{11}$ & 1.00 & 1.14 & 5.7 10$^{10}$ & 1.00 \\
A2911 & 8.41 10$^{11}$ & 0.88 & 1.25 & 0. & 1.00 \\
A2923 & 3.78 10$^{11}$ & 0.81 & 1.19 & 1.04 10$^{11}$ & 1.00 \\
A3111 & 8.01 10$^{11}$ & 0.75 & 1.24 & 0. & 1.00 \\
A3112 & 2.424 10$^{12}$ & 0.87 & 1.22 & 0. & 1.00 \\
A3122 & 6.65 10$^{11}$ & 0.81 & 1.16 & 9.8 10$^{10}$ & 1.00 \\
A3128 & 3.812 10$^{12}$ & 0.89 & 1.14 & 5.4 10$^{10}$ & 1.00 \\
A3141 & 1.311 10$^{12}$ & 1.00 & 1.44 & 0. & 1.00 \\
A3158 & 9.71 10$^{11}$ & 1.00 & 1.14 & 1.48 10$^{11}$ & 1.00 \\
A3202 & 2.71 10$^{11}$ & 0.66 & 1.19 & 1.0 10$^{10}$ & 1.00 \\
A3733 & 1.26 10$^{11}$ & 1.00 & 1.06 & 2.6 10$^{10}$ & 1.42 \\
A3764 & 3.23 10$^{11}$ & 0.82 & 1.23 & 3.3 10$^{10}$ & 1.21 \\
A3825 & 2.208 10$^{12}$ & 0.90 & 1.22 & 3.3 10$^{10}$ & 1.02 \\
A3827 & 8.96 10$^{11}$ & 1.00 & 1.38 & 4.4 10$^{10}$ & 1.02 \\
\hline	   
\normalsize
\end{tabular}
\end{flushleft}
\label{t-data2}
\end{table}

\begin{figure}
\vbox
{\psfig{file=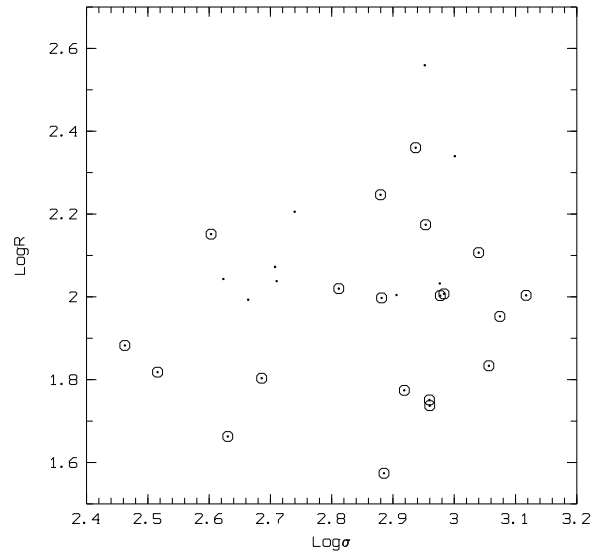,width=9.0cm,angle=270}}
\caption[]{$R$--$\sigma $  relation for King profile 
fits. The dotted  circles represent the more contrasted clusters 
	while the dots denote the less contrasted ones.}
\label{$L$sigma}
\end{figure}

All individual corrections are given in table 2; note however that the 
values of $C_1$ are given for information only, as we applied 
$\langle$ $C_1$ $\rangle $= 0.83 for all clusters.

Estimating the errors on the luminosities $L_{true}$ is not easy. From
the uncertainty in the slope of the luminosity function, we estimate
that $L_{true}$ is uncertain by 10~\%. To this we should probably add
other error sources but it is very difficult to make quantitative
estimates of those. Therefore we take the optimistic view of an error
of 10~\% in the luminosities.

To check the general consistency of our luminosities, we compare them
with those obtained by Carlberg et al. (1996). They found, in Gunn r, a 
median K-corrected luminosity for 16 distant ($z \approx 0.3$) clusters 
of 1.95 10$^{12}$h$^{-2}$ $L_{\odot _{r}}$, corresponding, to a value 
of 9.36 10$ ^{11}$ h$^{-2}$$L_{\odot _{b_j}}$, where we assumed 
$b_j-r=1.5$ and $b_j{\odot}-r_{\odot}=0.7$. Our mean cluster luminosity
is 7.65 10$^{11}$h$^{-2}$ $L_{\odot _{b_j}}$. Given that Carlberg et al. 
(1996) considered larger areas than we do (larger than 1.0 \Mpc vs. 0.9 
\Mpc), the agreement can be considered quite satisfactory.

\section{The Fundamental Plane}

As discussed in \S~2, we want to investigate possible 
$L$--$R$--$\sigma$ correlations for the galaxy clusters in our sample. 
In the process, we also look at possible $L$--$R$, $L$--$\sigma$ and 
$R$--$\sigma$ relations. Note that all fits were performed on the 
sample of the 20 most contrasted clusters (see below). The cluster A0168 
is the only one we have in common with S93.

\subsection{Fitting techniques}

\label{ss-fit}

We employ three fitting methods. First, we used the ESO MIDAS-package
which has an integrated fitting procedure that we used with and
without weights. The method consists of a classical least squares fit,
either unweighted, or weighted by the inverse of the square of the
error in velocity dispersion or characteristic radius.

We also used the MINUIT package to do a least squares fitting by
minimizing $\chi ^2$. This method has been developed to fit particle
trajectories and it is recognized to provide very good results. The
two MINUIT minimization methods are Simplex (Nelder et al. 1965) and
Migrad (Fletcher,1970). These two methods do not use derivatives.
Our strategy was to use Simplex to approach the final parameter values
and Migrad to solve for the parameters and estimate their errors. As
we show in paper VII, Simplex systematically underestimates the errors.
If Migrad does not converge, we take only the Simplex values without
errors. The strategy of using Migrad after Simplex in practice gives
good results and is commonly used, for example in the minimization
routines of the Greg numerical package.

\subsection{Results of the fitting}

\label{ss-res}

We performed the fitting of \sv vs. $R$ only on the 
subsample of the
20 more contrasted clusters in order to
avoid contaminating our sample with uncertain values of 
velocity
dispersion. The contrast is defined as the percentage of 
all galaxies
in the selected area which are cluster members. The 
number of
background galaxies was deduced from the background 
density calculated
by fitting the theoretical profiles (paper VII). The nine 
less 
contrasted clusters excluded from the fitting analysis 
are A0087, 
A0168, A1069, A2362, A2480, A2800, A2911, A3128 and 
A3825.

The clusters are indicated in Figs. 3, 4 and 5 by dots 
and dotted
circles, to distinguish the 9 less contrasted from the 20 
more 
contrasted clusters. 

\begin{figure*}
\vbox
{\psfig{file=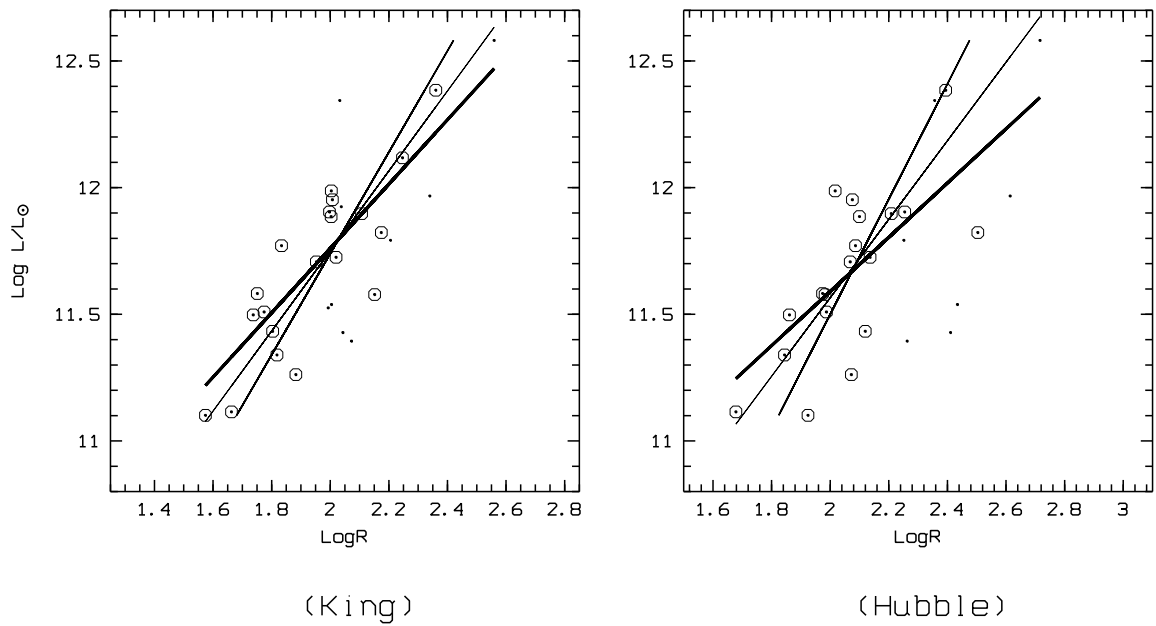,width=16.0cm,angle=270}}
\caption[]{$L$--$R$ relation for King (left) and Hubble 
(right) 
profiles (MIDWW calculations). The dotted circles 
represent the more
contrasted clusters and the dots are the less contrasted 
clusters. The 
heavy line is the $L$--$R$ relation, the
intermediate line the $R$--$L$ relation and the thin line 
is the
bisector relation, which has a slope of 1.58 for the King 
profile and
1.55 for the Hubble profile.}
\label{$L$$R$1}
\end{figure*}

We checked for the existence of a relation between \sv 
and R, where we
considered only King and Hubble radii, as discussed in 
Sec~3.  None of
the various fitting routines finds a relation between 
radius and
velocity dispersion (see e.g. Fig.~2, for King profile 
fits ), in
agreement with the conclusion of Girardi et al. (1996).

On the other hand, $R$ and $L$ {\em are} correlated. The 
fitting of
the $L$ vs. $R$ (and $R$ vs. $L$) relations was done using 
two MIDAS
regressions, viz. with (hereafter MIDWW) and without 
(hereafter
MIDWOUTW) weights on the fitted quantities. The fits were 
also made
using MINUIT (hereafter MWOUTW), without assigning any 
weight to the
data. Results of the fits are given in Table~3, where the 
parameters
$\alpha_1$, cst$_1$, $\alpha_2$ and cst$_2$ are defined 
by:
$(L/L_{\odot})$=$(R/kpc)^{\alpha_1}*10^{cst1}$ and
$(R/kpc)$=$(L/L_{\odot})$$^{\alpha_2}*10^{cst2}$. Note 
that only the
King and Hubble radii were considered. The weight that
we used here is the inverse of the square of the 
uncertainty in the
$R$-determination.
Fig.~3 shows the $R$ vs. $L$
scatter plots. In addition to the data, we also show the
fitted $L$--$R$ and $R$--$L$ relations, as well as the 
bisector line
which is the best estimator of a linear relation between 
two
parameters, according to Isobe et al. (1990).

\begin{table}
\caption[]{Fitted parameters for the two types of 
characteristic radius 
and the three fitting methods, for the $L$--$R$ 
relation.}
\begin{flushleft}
\small
\begin{tabular}{crcr}
\hline
\noalign{\smallskip}
parameters & MWOUTW & MIDWW & MIDWOUTW \\ 
\noalign{\smallskip}
\hline
\noalign{\smallskip}
$\alpha _1$ (King) & 1.38$\pm 0.24$ & 1.27$\pm 0.23$ & 
1.39$\pm 0.20$ \\
cst1 (King) & 8.98$\pm 0.24$ & 9.22$\pm 0.46$ & 8.98$\pm 
0.18$ \\ 
$\alpha _2$ (King) & 0.51$\pm 0.12$ & 0.50$\pm 0.09$ & 
0.51$\pm 0.08$ \\ 
cst2 (King) & -4.02$\pm 0.13$ & -3.87$\pm 0.10$ & 
-4.06$\pm 0.11$ \\  
$\alpha _1$ (Hubble) & 1.22$\pm 0.18$ & 1.07$\pm 0.27$ & 
1.22$\pm 0.29$ \\ 
cst1 (Hubble) & 9.12$\pm 0.21$ & 9.50$\pm 0.58$ & 
9.12$\pm 0.23$ \\ 
$\alpha _2$ (Hubble) & 0.42$\pm 0.09$ & 0.44$\pm 0.11$ & 
0.42$\pm 0.10$ \\ 
cst2 (Hubble) & -2.78$\pm 0.17$ & -3.11$\pm 0.33$ & 
-2.78$\pm 0.14$ \\ 
 \hline	   
\normalsize
\end{tabular}
\end{flushleft}
\label{t-data1}
\end{table}

\begin{figure}
\vbox
{\psfig{file=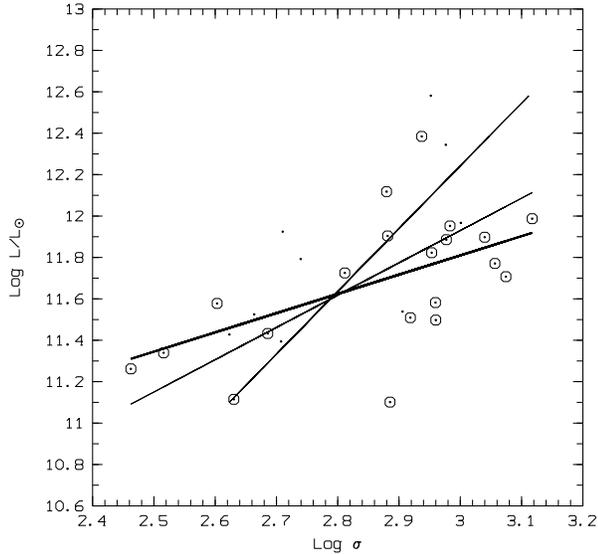,width=9.0cm,angle=270}}
\caption[]{The $L$--$\sigma $ relation (MIDWW 
calculations). The dotted
circles represent the more contrasted clusters and the 
dots the less
contrasted ones. The heavy line is the $L$--$\sigma$ 
relation, the
intermediate line the $\sigma$--$L$ relation and the thin 
line the
bisector relation which has a slope of 1.56.}
\label{$L$R}
\end{figure}

A correlation is also found between $\sigma$ and $L$. We 
fitted the 
parameters $\gamma _1$, cst$_1$, $\gamma _2$
and cst$_2$ in the relations 
$(L/L_{\odot})$=$\sigma^{\gamma
_1}*10^{cst1}$ and 
$\sigma$=$(L/L_{\odot})$$^{\gamma_2}*10^{cst2}$
(see Tab.~4 and Fig.~4), with $\sigma$ in km s$^{-1}$. 
The weight that
we used here is the inverse of the square of the 
uncertainty in the
$\sigma$-determination.

\begin{table}
\caption[]{Fitted parameters for the two types of 
characteristic radius 
and the three fitting methods, for the $L$--$\sigma $ 
relation. }
\begin{flushleft}
\small
\begin{tabular}{crcr}
\hline
\noalign{\smallskip}
parameters & MWOUTW & MIDWW & MIDWOUTW \\ 
\noalign{\smallskip}
\hline
\noalign{\smallskip}
$\gamma _1$ & 1.04$\pm 0.32$ & 0.93$\pm 0.27$ & 1.04$\pm 
0.33$ \\ 
cst1 & 8.69$\pm 0.29$ & 9.02$\pm 0.80$ & 8.67$\pm 0.28$ 
\\ 
$\gamma _2$ & 0.33$\pm 0.13$ & 0.33$\pm 0.11$ & 0.33$\pm 
0.11$ \\ 
cst2 & -1.04$\pm 0.15$ & -1.04$\pm 1.20$ & -1.04$\pm 
0.16$ \\ 
\hline 
\normalsize
\end{tabular}
\end{flushleft}
\label{t-data1}
\end{table}

Finally we fitted the parameters $\alpha $ and $\beta $ 
in the
relation $(L/L_{\odot})$=$R^\alpha \sigma^\beta 
*10^{cst}$, with $R$
in kpc and $\sigma$ in km s$^{-1}$, for both the King and 
Hubble radii
(see Fig.~5 and Tab.~5). We have made a fit which 
minimizes the
r.m.s. deviation in the $L$ direction. The weight used in 
the MIDWW
regression is the inverse of the square of the error in 
$R$. From
Fig.~5 one can see that the less contrasted clusters have 
a slightly
larger dispersion around the fit than the more contrasted 
ones. We have
verified that the parameters and the dispersion 
around the fit do 
not vary significantly if we slightly change the area 
in which we 
calculate the luminosity. In addition, the dispersion in the 
$L$--$R$--$\sigma$ relation for the King radius with a 
luminosity calculated within 4 King radii instead of 5, is smaller 
than when the luminosity is calculated within 4 Hubble radii.

\begin{table}
\caption[]{Fitted parameters for the two types of 
characteristic radius
and the three fitting methods, for the $L$--$R$--$\sigma 
$ relation.}
\begin{flushleft}
\small
\begin{tabular}{crcr}
\hline
\noalign{\smallskip}
parameters & MWOUTW & MIDWW & MIDWOUTW \\ 
\noalign{\smallskip}
\hline
\noalign{\smallskip}
$\alpha $ (King) & 1.24$\pm 0.12$ & 1.19$\pm 0.14$ & 
1.25$\pm 0.13$ \\ 
$\beta $ (King) & 0.78$\pm 0.15$ & 0.91$\pm 0.16$ & 
0.78$\pm 0.14$ \\ 
cst (King) & 7.00$\pm 0.13$ & 6.74$\pm 0.51$ & 7.00$\pm 
0.11$ \\ 
$\alpha $ (Hubble) & 0.98$\pm 0.09$ & 0.87$\pm 0.26$ & 
0.98$\pm 0.26$ \\ 
$\beta $ (Hubble) & 0.71$\pm 0.11$ & 0.70$\pm 0.31$ & 
0.71$\pm 0.26$ \\
cst (Hubble) & 7.60$\pm 0.25$ & 7.92$\pm 0.86$ & 7.60$\pm 
0.20$ \\ 
\hline
\normalsize
\end{tabular}
\end{flushleft}
\label{t-data1}
\end{table}

Following J$\o$rgensen et al. (1996), we have also 
minimized the r.m.s.
deviations in the two other directions (Table~6). We ran 
only a MIDWW
fit for the King profile. We find no significant 
variations with
minimization direction. This supports the reliability of 
our fitting
results.

The differences between the values of the parameters 
obtained with
different methods are within the fitting errors. There is 
a slight
apparent inconsistency in the results of the 2-parameter 
fits  when 
minimized in the two parameter directions. For example, 
the slope $\gamma_1$
and the slope $\gamma_2$ in the relation between 
luminosity and
velocity dispersion certainly do not obey $\gamma_1 
\approx \frac
1{\gamma_2}$. However, it is well known that one has
$\gamma_1$ = $\frac 1{\gamma_2}$ only if the correlation 
coefficient
between $L$ and $\sigma$ is 1. In other words, one must 
have the same
covariance between $L$ and $\sigma$ as between $\sigma$ 
and $L$ which
is clearly not the case. On the other hand, the 
coefficients of the
$L$--$R$--$\sigma$ relations do not depend significantly 
on the
minimization direction. The dispersion in the 3-parameter 
relation is
smaller than in the two 2-parameter relations, and the 
correlation
coefficient is close to unity.

\begin{table}
\caption[]{Fitted parameters for the cluster 
$L$--$R$--$\sigma $ 
           relation, for the three minimization 
directions, using the
           King radius}
\begin{flushleft}
\small
\begin{tabular}{ccc}
\hline
\noalign{\smallskip}
direction & \multicolumn{2}{c}{($L \propto R$$^\alpha 
\sigma ^\beta$)} \\ 
          &  $\alpha$ & $\beta$ \\ 
\noalign{\smallskip}
\hline
\noalign{\smallskip}
$L$ & 1.19$\pm 0.14$ & 0.91$\pm 0.16$ \\ 
$R$ & 1.43$\pm 0.14$ & 0.87$\pm 0.32$ \\ 
$\sigma$ & 1.37$\pm 0.21$ & 1.15$\pm 0.38$ \\  
\hline
\normalsize
\end{tabular}
\end{flushleft}
\label{t-data1}
\end{table}

\begin{figure*}
\vbox
{\psfig{file=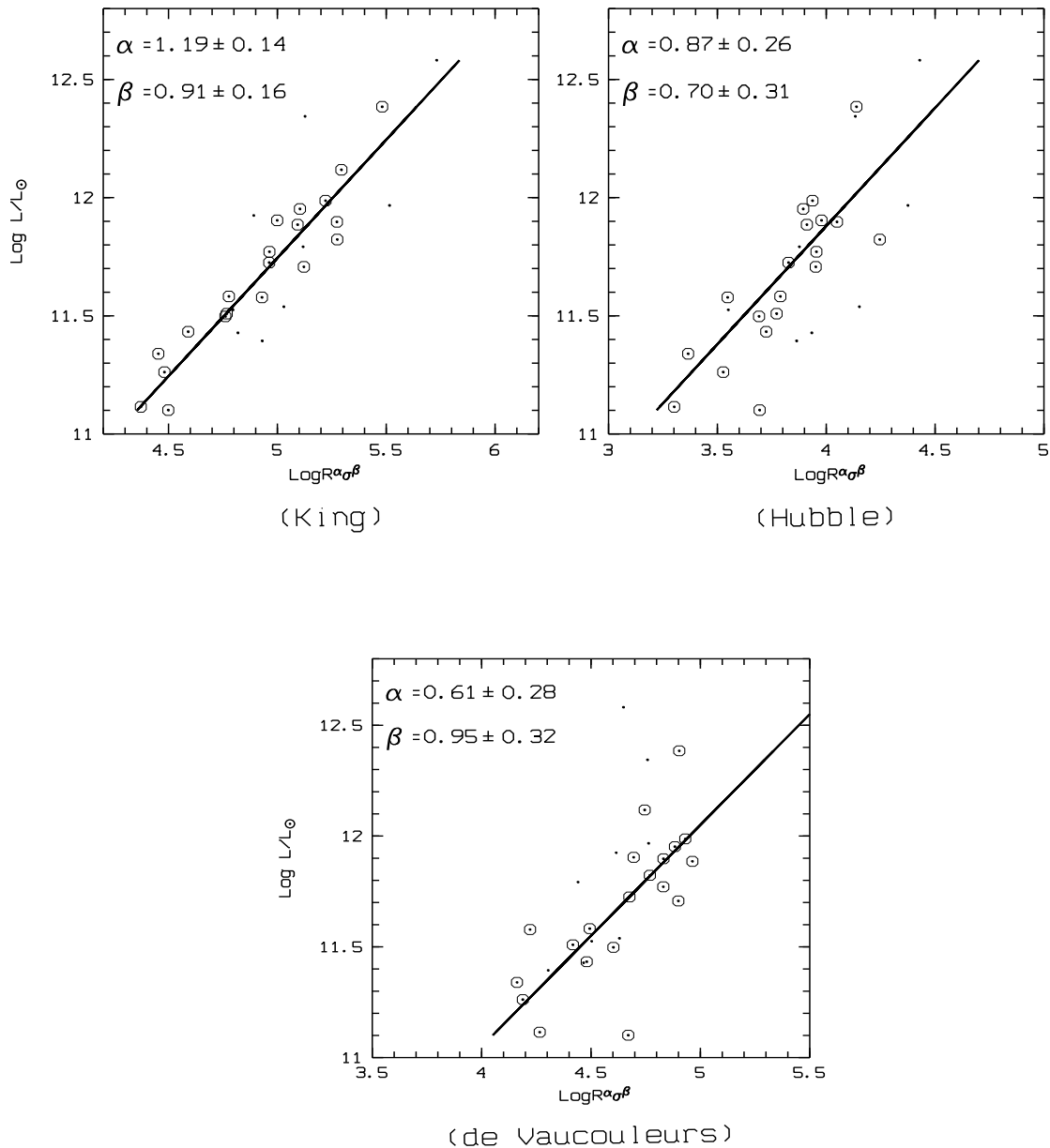,width=16.0cm,height=18.0cm,angle=180.}}
\caption[]{$L$--$R$--$\sigma $ relation for King, Hubble 
and 
de Vaucouleurs profiles (MIDWW calculations). The dotted
circles represent the more contrasted clusters and the 
dots are the 
less contrasted ones. The abscissa is equal to the 
logarithm of 
$R^\alpha \sigma^\beta$.
}
\label{$L$$R$2}
\end{figure*}

\section{Discussion}
\label{s-disc}

We have confirmed the result of Schaeffer et al. (1993) that clusters 
of galaxies populate a Fundamental Plane (FP). The orientation 
of the FP of
clusters can be determined reasonably well from our ENACS 
data, but
the details depend somewhat on the `direction' in which 
we minimize
the r.m.s. deviations from the FP. In particular, we find 
a slightly
steeper dependence of $L$ on $\sigma$ if we minimize in 
the
$\sigma$-direction, although the difference is not 
significant. 
In Fig.~5 we show what is essentially a sideways 
view of the FP by
projecting the observations in a direction parallel to 
the FP, onto a 
plane that is orthogonal to the FP. The reality of the 
inclination of the 
FP in the (log $L$, log $R$, log $\sigma$)-space is 
evident already in 
Figs.~2, ~3 and ~4, which represent projections along the 
coordinate axes. 
In Fig.~6 we show an attempt at a 3-dimensional 
visualisation of the FP 
in the (log $L$, log $R$, log $\sigma$)-space. This figure is 
for King profile fits. 

The comparison with the FP parameters obtained by S93, 
for a different
sample of clusters, requires some care because S93 used a 
de
Vaucouleurs profile fit. Therefore, we have
also repeated our analysis for de Vaucouleurs-profile 
fits for 
our clusters. However, we stress once again that this
profile provides a fit to the data that is inferior to 
that
of a King or Hubble profile (see also paper VII).

In Tab.~7 we give the FP parameters obtained from the 
ENACS clusters
for the 3 types of profile just mentioned. Although the 
effect is not highly significant in view of the errors, there 
is a tendency for 
the L-dependence on $R$ to decrease from King to Hubble 
to de Vaucouleurs
profile. However, there is no apparent change of the 
dependence of $L$
on $\sigma$ along the same sequence. We also show in the 
same table
the FP parameters derived by S93. There is a hint that 
the relation
between $L$ and \sv (for the de Vaucouleurs profile) may 
be steeper in
S93 than it is in our data, but the difference of about 30\% 
in $\beta$ is
within one sigma. It is therefore evident that the
present results qualitatively agree well with those of 
S93, although there are some quantitative differences. This
provides additional evidence of the reality of the galaxy 
cluster FP, because both the
data-samples and the methods used are different.

We note in passing that West et al. (1989) derived an 
$L$-$R$
relation, using the same data as S93, that also agrees 
quite well with
our result.

For clusters in virial equilibrium, which have a fixed 
value of
$M$/$L$, one expects $L \propto R \sigma ^2 $, i.e. 
$\alpha = 1.0$ and
$\beta = 2.0$.  Clearly, our $\beta$-value is different 
from the
virial prediction. In this respect, there is a similarity 
between
galaxy clusters and elliptical galaxies. The deviation 
from the pure
virial relation, as apparent from the cluster FP, may not
seem very large but it is quite significant. This is apparent
from the orthogonal dispersion of the 20 clusters around the virial 
relation of 0.12, which is more than two times larger than the
dispersion around
our best-fitting FP of 0.05 (see also Table 8). For the sake of the 
present
discussion, we will therefore assume that the FP-fit to 
our data
provides a considerably better description of the
$L-R-\sigma$-relation of galaxy clusters than does the 
virial
relation.

\begin{figure*}
\vbox
{\psfig{file=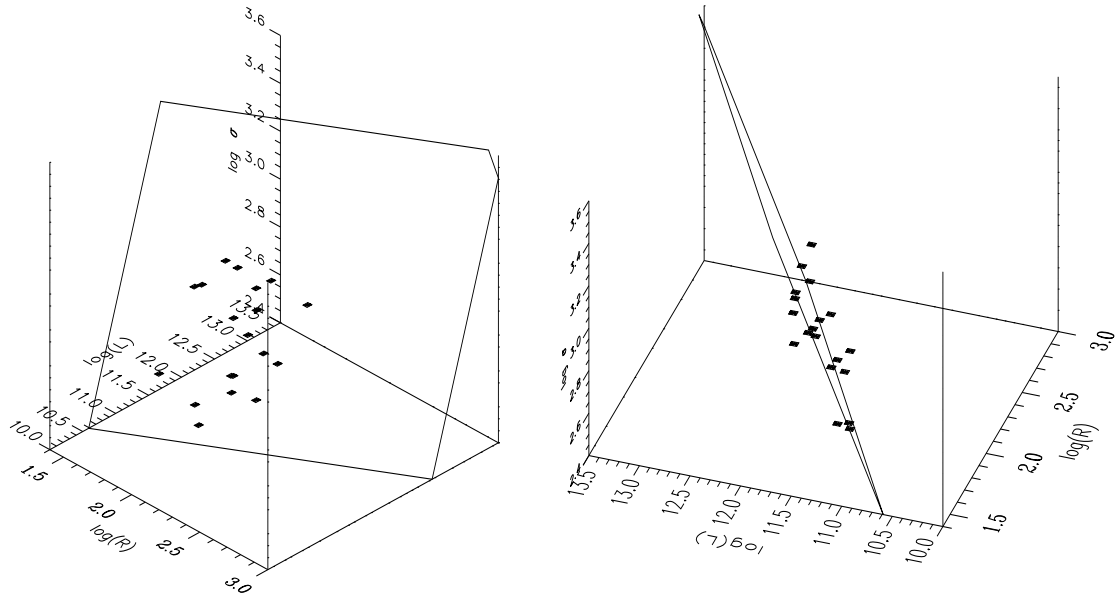,width=16.0cm,angle=270.}}
\caption[]{Two projections of the Fundamental Plane of clusters, in 
the (log $L$, log $\sigma$, log $R$)-space, which show the distribution 
of clusters in the plane (left) and the dispersion around the plane 
(right).}
\label{3D}
\end{figure*}

The virial theorem can be expressed as follows (Kormendy 
$\&$
Djorgovski, 1989):

$L$ = $S$ $R_d$ $\frac{L}{M}$ $R$ $\sigma_v^2$

\noindent In this equation, $S$ is a parameter related to 
the internal 
structure of the system, and $R_d$ is the ratio of the 
potential to
the kinetic energy of the system, and it measures the 
degree of
relaxation of the system. The deviations of the 
($\alpha$,$\beta$)
coefficients from the (1,2) virial prediction could be 
the result of a
non-constant value for the product of $S$, $R_d$ and 
$M$/$L$ for
the different clusters. 

As discussed in S93, the case of non-constant $M$/$L$ has 
a
special solution if $\frac{M}{L} \propto L^\varepsilon$; 
then one has
$\beta = 2 \alpha$. For the FP fit in S93, with $\alpha = 
0.89 \pm
0.15$ and $\beta = 1.28 \pm 0.22$, this special case 
could indeed
apply. If we take our result for de Vaucouleurs fits to 
the density
profiles ($\alpha = 0.61 \pm 0.28$ and $\beta = 0.95 \pm 
0.32$) we
would again conclude that the data are consistent with 
this special
case. However, we must stress again that the de 
Vaucouleurs profile
fits are notably worse than the King profile fits. 
Comparing therefore
the values $\alpha = 1.19 \pm 0.20$ and $\beta = 0.91 \pm 
0.26$ that
we found for the King profile fits, we conclude that 
$M$/$L$ is likely
not to have a simple power-law dependence on $L$.

This would then imply that probably also different $S$ 
and $R_d$ are
needed for different clusters to explain the deviation of 
the cluster
FP from the virial relation. Another way to summarize the 
situation
could be to say that the structure of clusters is such 
that the simple
equilibrium density laws {\em either} do not fit the data 
very well
(de Vaucouleurs), while $M$/$L$ can be assumed 
proportional to
$L$ or $M$, {\em or} they fit quite nicely (King) but 
then it is
unlikely that $M$/$L$ has a simple power-law dependence 
on $M$.

We can set an upper limit to the variation of $M$/$L$ 
among clusters,
if we assume the virialization state and internal 
structure of all
clusters to be identical. In that case $M$/$L$ $\propto 
R^{1-\alpha}
\sigma_v^{2-\beta}$, or approximately $M$/$L$ $\propto 
\sigma_v^{2-\beta}$ because $\alpha \approx$ 1. I.e., 
systems with 
larger \sv have larger $M$/$L$. The $M$/$L$ ratio of rich 
clusters
would then be expected to vary at most a factor of 2 to 
3, given the
observed \sv-range of rich galaxy clusters (see paper 
II).

If the $M$/$L$ ratio of rich clusters indeed is more or less
proportional to $\sigma_v$, this could fundamentally affect the 
determination of the density parameter of the Universe from 
clusters which
are acting as gravitational lenses. These clusters are 
generally among
the most massive ones so they have a large $\sigma_v$. 
Therefore they
could have atypically high $M$/$L$ ratios which would 
lead to an
overestimation of the density parameter (e.g. Bonnet et 
al. 1994).

\begin{table*}
\caption[]{Fundamental Plane parameters for galaxy 
clusters and for 
           elliptical galaxies.}
\begin{flushleft}
\small
\begin{tabular}{cccc}
\noalign{\smallskip}
\hline
\noalign{\smallskip}
Authors & Type of system & $\alpha $ & $\beta $ \\ \hline
present paper & galaxy clusters (King profile) & 1.19$\pm 
0.14$ & 0.91$
\pm 0.16$ \\
present paper & galaxy clusters (Hubble profile) & 
0.87$\pm 0.26$ & 0.70$
\pm 0.31$ \\ 
present paper & galaxy clusters (de Vaucouleurs profile) 
& 0.61$\pm 0.28$
 & 0.95$\pm 0.32$ \\
Schaeffer et al. & galaxy clusters (de Vaucouleurs 
profile) & 0.89$\pm $
0.15 & 1.28$\pm $0.22 \\ 
Djorgovski \& Davis & Virgo ellipticals & 0.89 & 1.54 \\ 
Bender et al. & cluster ellipticals & 0.82 & 1.64 \\ 
Guzman et al. & cluster ellipticals & 0.72 & 1.44 \\ 
Pahre et al. (1995) & cluster ellipticals & 0.73$\pm 
0.07$ & 1.82$\pm 0.14$ \\
J$\o$rgensen et al. & Coma ellipticals & 0.78$\pm 0.05$  
& 1.51$\pm 0.09$ \\ 
\hline
\normalsize
\end{tabular}
\end{flushleft}
\label{t-data1}
\end{table*}

It is of interest to compare the cluster FP with that of 
elliptical
galaxies. The results of Djorgovski \& Davis (1987), 
Bender et
al. (1992), Guzman et al. (1993), Pahre et al. (1995) and 
J$\o$rgensen
et al. (1996), are listed in Tab.~7. As was done by 
J$\o$rgensen et
al. we translated, if necessary, the published FP 
solution to the form
we used ($L \propto R$$^\alpha \sigma ^\beta$) by using 
the relation: $<~I_e~> \propto L/{r_e}^2$. Of course, this is not 
exactly identical as we determined $L$, $\sigma$ and $R$ 
independently, whereas $r_e$ and $<~I_e~>$ were 
derived from the same luminosity profile, and are therefore 
somewhat correlated. We have also 
used the global, overall velocity dispersion, while for 
ellipticals it is 
the central velocity dispersion that is used. However, the 
velocity dispersion profiles of galaxy clusters are in general 
quite flat (see e.g. den Hartog and Katgert 1996). Note that 
the result of
Pahre et al. refers to the near-infrared, while the other 
results
refer to the optical.

The $\alpha$-values found for ellipticals by the various 
authors, in
the optical and near-infrared, appear quite consistent. 
The
$\beta$-values found by the different groups are also 
quite similar,
although the value found in the near-infrared (Pahre et 
al., 1995) is 
slightly high compared to the optical values. This is even 
more evident in the 
recent result by Pahre et al. (1996) 
who find, again in the near-infrared, $\alpha$ = 0.67 and 
$\beta$ =
2.21.

Comparing the various determinations of $\alpha$ and 
$\beta$ in
Tab.~7, we cannot be certain that the values of $\alpha$ 
are
significantly different for ellipticals and for galaxy 
clusters. This
does not mean that they are identical. As we said before, 
the
agreement in $\alpha$ is good for de Vaucouleurs fits to 
the cluster
profiles, but much less (about 2 $\sigma$) for the King 
profile that 
describes our cluster observations much better. The 
situation appears much 
clearer for $\beta$: independently of the type of profile 
used, the
$\beta$-value for the clusters appears significantly 
lower ($\ge$ 2$\sigma$) 
than that of the ellipticals and much more when compared 
to the result described by Pahre (1996).

With caveats of the previous paragraphs in mind, it seems
safe to conclude from Tab.~7 that the FP's of 
clusters of
galaxies and of elliptical galaxies have different 
orientations. It is therefore quite unlikely that the
two types of system share a common, universal FP, and we do not
think that our data support the conclusion to this effect in S93. 
If our result is confirmed, this will have very important
implications for the formation of gravitationally bound 
systems on
different scales. A more accurate determination of the 
cluster FP is
needed in order to quantify the differences between the 
FP's of galaxy
clusters and of elliptical galaxies more accurately 
before one can
start to investigate the implications in more detail.

For the moment, we can only speculate about the possible 
implications.
Taking the values in Tab.~7 at face value, we are struck 
by the fact
that the deviations from the virial prediction seem to be 
different
for galaxy clusters and ellipticals. For the ellipticals 
it seems that
the value of $\beta$ may be quite close to the virial 
value
(especially in the near-infrared). Or, if one gives more 
weight to the
optical data, the ellipticals at least seem to obey the 
relation
$\beta = 2 \alpha$ quite well. None of that is true for 
the clusters,
for which it seems quite likely that $\alpha$ agrees with 
the virial
prediction (which probably is not the case for ellipticals 
!), while the
value of $\beta$ seems definitely at odds with the virial 
expectation.

Therefore, it is possible (if not likely) that the deviations 
from the virial
prediction have quite different origins for ellipticals 
and for
clusters. For ellipticals the deviations may well be due 
to
non-homology and non-constant M/L ratio's. For clusters 
the deviations
may be due primarily to their relative youth or, in other 
words: to an
absence of real equilibrium. Even though we selected the 
most regular
and apparently relaxed clusters for the present analysis, 
it is
unlikely that they have really attained equilibrium 
except in their
very centres (see e.g. den Hartog and Katgert 1996, and 
paper III). If
this is true, it is not immediately clear why we should 
still find a
well-defined FP for these clusters which probably are not 
fully
relaxed.

The orthogonal dispersion of the 20 well-contrasted 
clusters around
the FP that they define is 0.05. This is supported by 
visual
inspection of the upper left-hand panel of Fig.~5. However, 
inspection of
that same figure also reminds one of the fact that the 
other 10
clusters are distributed less narrowly around the FP. For 
the moment
we will assume that these less contrasted and less 
regular clusters
are probably not as relaxed as the other 20 clusters, so 
that one
cannot expect them to define an FP as narrowly as the 
more contrasted
and more regular clusters. It is interesting to observe 
in Fig.~5 that
the dispersion around the FP is indeed smallest for the 
King profile,
somewhat larger for the Hubble profile and largest for 
the de
Vaucouleurs profile. This supports in a rather 
independent way the
conclusion in paper VII that de Vaucouleurs profiles do 
not provide
good fits to the galaxy distributions in clusters.

We have tried to estimate how much of the scatter in the 
FP is due to
measurement errors and how much to the intrinsic width of 
the FP. To
that end we have done the following experiment. We have 
assumed the
solution for the FP, with $\alpha = 1.19$ and $\beta$ = 
0.91, to be
correct. Therefore, we have taken the orthogonal 
projections of the
observed points onto that FP as starting points for the 
following
simulations. For the set of 20 points {\em in the FP} 
obtained thus,
we simulated 500 artificial sets in which uncorrelated 
errors are added
in $L$, $R$ and $\sigma$ to each of the 20 points, 
according to
Gaussian distributions with dispersions as given in 
Tab.~1 and section 
3.3. We assumed an error of 10\% for $L$. 

For each of the 500 sets of 20 artificial points, the FP 
was solved in
exactly the same way as it was done for the observations 
with the 
MIDWW method. As a result
we obtain 500 pairs ($\alpha$,$\beta$). The distributions 
of the 500
values of $\alpha$ and $\beta$ yield average values and 
dispersions of
$1.17 \pm 0.06$ and $0.96 \pm 0.10$ for $\alpha$ and 
$\beta$
respectively. These values must be compared to the values 
and their
estimated errors from the FP fit to the observations of 
$1.19 \pm
0.14$ and $0.91 \pm 0.16$. The estimated errors in the 
latter values
are calculated from the assumed errors in $L$, $R$ and 
$\sigma$. 

The average values of $\alpha$ and $\beta$ in the 500 
artificial sets
are not identical to the input values (the most probable 
values found
in the FP fit to the data) but they are quite close. On 
the other
hand, the dispersions in the $\alpha$ and $\beta$ values 
of the 500
artificial sets are significantly smaller than the 
estimated
uncertainties found in the FP fit to the data. We 
interpret this as
evidence that the apparent width of the FP of 0.05 (for 
the King
profile fits) has a important contribution from the 
intrinsic width of
the FP, i.e. that it is not exclusively due to 
measurement errors.

It is not trivial to estimate how much of the apparent 
width of the
observed FP is intrinsic. One can get some idea from the 
average
dispersion around the 500 artificial datasets of 20 
`observed' points,
generated as described above. The average dispersion 
around these 500
individual FP's is 0.06, i.e. larger than the value 0.05 
for the
'optimum' FP, as it should be. This provides a crude 
estimate of the
noise contribution to the width of the FP of about 
0.03--0.04, which
then implies a similar range of values for the intrinsic 
width of the
FP.

In Table~8 we show the dispersions around the other fits 
for all
relations that we studied in this paper, and with the 
present
dataset. Clearly, the dispersions around the FP are 
smaller than those
around the $L$--$R$ and $L$--$\sigma$-relations, for all 
types of
profile fitted. The differences in the dispersions around 
the
$L$--$R$--$\sigma$-relations are striking. They show that 
the King
profile not only provides the best description of the 
central, regular
parts of galaxy clusters, but that it also provides a
significantly narrower FP than any other of the currently 
popular
profiles.

\begin{table}
\caption[]{Orthogonal dispersions around the three fitted 
relations for 
the MIDWW coefficients. 
and
}
\begin{flushleft}
\small
\begin{tabular}{cccc}
\noalign{\smallskip}
\hline
\noalign{\smallskip}
    & $L$--$R$--$\sigma $ & $L$--$R$ & $L$--$\sigma $  \\ 
\hline
King profile                      & 0.05 & 0.08 & 0.14 \\ 
Hubble profile                    & 0.09 & 0.11 & 0.14 \\
Virial coeff. $\&$ King profile     & 0.12 & 0.12 & 0.17 
\\
de Vaucouleurs profile            & 0.14 & 0.14 & 0.14 \\
\hline
\normalsize
\end{tabular}
\end{flushleft}
\label{t-data1}
\end{table}

One might think that the large dispersion in the $L$-\sv 
relation
could be partly due to systematic errors in \sv that are 
due to
contaminating groups of galaxies projected onto the 
cluster core. Such
galaxies could be falling into the cluster and would not 
be virialized
in the cluster potential. However, that is not likely to 
be an
important effect, because the velocity dispersions vary 
by only a few
percent if we exclude the emission-line galaxies, which 
are the ones
suspected to be on fairly radial, infalling orbits (see, 
e.g., paper
III). Apparently our selection of the most regular (i.e. 
probably
virialized) cluster cores, has already minimized this 
effect. The fact
that the scatter around the FP appears to increase when 
lower-contrast
and less regular clusters are included (see Fig.~5) could 
partly be due to 
such contaminating galaxies. 

On the other hand, the increase of the scatter when 
lower-contrast and
less regular clusters are included could equally well be 
the result of
the fact that these are apparently less well relaxed. The 
larger
dispersion around the FP for the latter clusters is not 
very likely to
be due to relatively larger formal uncertainties in $L$, 
$R$ or
$\sigma$, as inspection of Tab.1 will show.

We conclude that a significant part of the dispersion of 
clusters
around the FP is intrinsic. As we discussed earlier, 
there are many physical effects that may be responsible for the 
intrinsic width
of the FP: structural differences among clusters, 
differences in
virialization state, and peculiar velocities with 
respect to a
uniform Hubble flow.

Deviations from a pure Hubble flow result in errors in 
the distance of
clusters, $\delta D$, which translate into an error 
$\delta L$/$L$ = 2
$\delta D/D$ and an error $\delta R$/$R$ = $\delta D/D$. 
Therefore 
the orthogonal intrinsic
dispersion around the FP of 0.03--0.04 translates into a 
dispersion in
the $L$-direction of about 0.05. As our clusters are at 
an average cz of 
20000 \ks, it is thus unlikely that they
have peculiar velocities that greatly exceed 1000 \ks, as 
those would
induce a larger dispersion around the FP than we observe. 
This is in
agreement with S93 and with the recent estimates based on 
Tully-Fisher
galaxy distances by Bahcall \& Oh (1996).

\section{Conclusions}

We have re-examined the evidence for the existence of a 
Fundamental
Plane (FP) for galaxy clusters, using the new data from 
the ESO Nearby
Abell Cluster Survey (for velocity dispersions and 
correction for
field contamination), and the Cosmos Galaxy
Catalogue (for the characteristic radii of the galaxy 
distributions
and the luminosities). We have derived the luminosities, 
radii and
velocity dispersions for our dataset of 29 rich clusters 
in a
homogeneous manner, and we have studied the correlations 
between these
parameters. Our clusters are found to define a quite
narrow FP, which
confirms the result obtained earlier by S93.

Our result is qualitatively similar to that of S93
although there are quantitative differences that are 
partly due to
the use of different profiles to describe the projected 
galaxy
distribution. On the other hand, our result appears to 
indicate that
the FP for galaxy clusters is significantly different 
from that for
elliptical galaxies. However, more work is needed to 
establish
accurately the qualitative difference between the two, 
which is a
prerequisite for a more detailed understanding of the physical 
implications of the
difference.

The FP that we find for our galaxy clusters differs from 
the virial
prediction, primarily in the sense that the cluster 
luminosity appears
to vary linearly, rather than
quadratically, with velocity dispersion. There are 
several possible
effects that could be responsible for that. E.g., cluster 
cores may
have a variety of dynamical structures, or $M$/$L$ may not be constant, 
or they may not all be fully virialized, 
(or combinations of these). Assuming that non-constant $M$/$L$ is 
the only reason for
the difference between the observed FP and the virial 
prediction, we
estimate that the $M$/$L$ ratio of rich clusters varies 
by at most a
factor of 2 to 3.

We conclude that a significant part of the observed 
scatter around the
FP is likely to be intrinsic. However, there may also 
be a
contribution from distance errors that are induced by 
peculiar
velocities of clusters with respect to the Hubble flow. 
Assuming that
all the intrinsic scatter in the FP is due to deviations 
from a pure
Hubble flow, we can set an upper limit to the typical 
cluster peculiar
velocities of about 1000 \ks, in agreement with other, 
independent
results.

\begin{acknowledgements}

{CA, AM, AB and PK acknowledge financial support from the 
French GDR
Cosmologie and INSU. AM also acknowledges financial 
support by Leiden 
Observatory. CA thanks also A. Maucherat, M. Bout and J.M. Deltorn
for helpful discussions.}

\end{acknowledgements}

\vfill
\end{document}